\documentclass[mathematics,article,accept,pdftex,moreauthors]{Definitions/mdpi} 
\usepackage{algorithm}
\usepackage{algpseudocode}
\usepackage{color}
\pdfoutput=1
\firstpage{1} 
\pubvolume{1}
\issuenum{1}
\articlenumber{0}
\pubyear{2022}
\copyrightyear{2022}
\datereceived{} 
\dateaccepted{} 
\datepublished{} 
\hreflink{https://doi.org/} % If needed use \linebreak
\Title{Two-dimensional correlation
analysis of periodicity in noisy series: Case of  VLF signal amplitude variations  in the time vicinity of  an earthquake}

\TitleCitation{2D Hybrid method:Case of VLF signal amplitude variations  in the time vicinity of  an earthquake}

\Author{Andjelka B. Kova\v cevi\'c $^{1,2}$*\orcidA{}, Aleksandra Nina $^{3}$\orcidB{} , Luka \v C. Popovi\'c $^{1,2,4}$\orcidC{} and
Milan Radovanovi{\'c} $^{5,6}$\orcidD{}}

\AuthorNames{Andjelka B. Kova\v cevi\'c, Aleksandra Nina, Luka \v C. Popovi\'c, Milan Radovanovi\'c}

\AuthorCitation{Kova\v cevi\'c, A.B.; Nina, A.; Popovi\'c, L.\v C., Radovanovi\'c, M.}
\address{%
$^{1}$ \quad University of Belgrade-Faculty of Mathematics, Department of astronomy,
Studentski trg 16
Belgrade, Serbia; andjelka@matf.bg.ac.rs\\
$^{2}$ \quad PIFI Research Fellow, Key Laboratory for Particle Astrophysics, Institute of High Energy Physics, Chinese Academy of Sciences,19B Yuquan Road, 100049 Beijing, China\\
$^{3}$ \quad Institute of Physics Belgrade, University of Belgrade, Pregrevica 118 Belgrade, Belgrade, Serbia\\
$^{4}$\quad Astronomical Observatory, Volgina 7, 11160 Belgrade Serbia; lpopovic@aob.rs\\
$^{5}$\quad Geographical Institute Jovan Cviji{\' c} SASA, Belgrade 11000, Serbia\\
$^{6}$\quad South Ural State University, Institute of Sports, Tourism and Service, 454080 Chelyabinsk, Russia
}

\corres{Correspondence: andjelka@matf.bg.ac.rs (A.K.)}

\abstract{
Extraction of information in the form of oscillations from noisy data of natural phenomena such as sounds, earthquakes, ionospheric and brain activity, and various emissions from cosmic objects is extremely difficult. As a method for finding periodicity in such challenging data sets, the 2D Hybrid approach, which employs wavelets, is presented.
Our technique produces a wavelet transform correlation intensity contour map for two (or one) time series on a period plane defined by two independent period axes. Notably, by spreading peaks across the second dimension, our method improves apparent resolution of detected oscillations in the period plane and identifies the direction of signal changes using correlation coefficients.
We demonstrate the performance of the 2D Hybrid technique on a very low frequency (VLF) signal emitted in Italy and recorded in Serbia in time vicinity of the occurrence of an earthquake on November 3, 2010, near Kraljevo, Serbia. 
We identified a distinct signal in the range 120-130 s that appears only in association with the considered earthquake. 
Other wavelets, such as Superlets, which may detect fast transient oscillations, will be employed in the future analysis.
}


\begin{document}

%%%%%%%%%%%%%%%%%%%%%%%%%%%%%%%%%%%%%%%%%%
%\setcounter{section}{-1} %% Remove this when starting to work on the template.
%\section{How to Use this Template}

%The template details the sections that can be used in a manuscript. Note that the order and names of article sections may differ from the requirements of the journal (e.g., the positioning of the Materials and Methods section). Please check the instructions on the authors' page of the journal to verify the correct order and names. For any questions, please contact the editorial office of the journal or support@mdpi.com. For LaTeX-related questions please contact latex@mdpi.com.%\endnote{This is an endnote.} % To use endnotes, please un-comment \printendnotes below (before References). Only journal Laws uses \footnote.

% The order of the section titles is: Introduction, Materials and Methods, Results, Discussion, Conclusions for these journals: aerospace,algorithms,antibodies,antioxidants,atmosphere,axioms,biomedicines,carbon,crystals,designs,diagnostics,environments,fermentation,fluids,forests,fractalfract,informatics,information,inventions,jfmk,jrfm,lubricants,neonatalscreening,neuroglia,particles,pharmaceutics,polymers,processes,technologies,viruses,vision

\section{Introduction}

 {Monitoring of different parts of the Earth and space collects data whose analyses can provide numerous important pieces of information for both scientific research and practical applications. One of the most important applications of various forms of monitoring is in the field of natural disaster prediction. However, in many cases, the possibilities and reliability of the corresponding predictions are still in the research phase. One of these examples is the application of data obtained in the monitoring of the lower ionosphere with very low frequency (VLF) signals to the prediction of earthquakes. These signals are emitted by worldwide located transmitters and propagate in the so-called "Earth-ionosphere waveguide", while numerous receivers record the signal amplitude and phase at their locations. Variations in the characteristics of the recorded signal enable the indirect detection of numerous phenomena, among which are those related to natural disasters. Among others, a large number of studies based on the analysis of VLF signals investigate the possibility of the existence of earthquake precursors in the form of ionospheric perturbations. Those precursors are primarily associated with changes in the VLF signal amplitude and/or phase  \citep{bia06,bia01a,bia01b,roz04,zha20} and  the amplitude minimum time shift during solar terminator periods (the so-called "terminator time") \citep{hay96,mau16,mol98,yam07,yos08}. In addition, the most recent research suggests that there are reductions in the noise of the VLF signal amplitude and phase a few tens of minutes before the observed type of disaster \citep{nin20a,nin21b,nin22b}. Although various data processing procedures have been applied in previous studies, there is still no way to reliably predict an earthquake. For this reason, the application of new models in VLF signal processing is essential.}

Signals are  similar to quantum systems due to wave-particle duality. 
 The {scientist who first noticed} this and prove uncertainty principle for signals was Gabor \cite{Gab46}, who did so by applying to arbitrary signals the same mathematical apparatus that was employed in the Heisenberg-Weyl \cite{doi.org/10.1007/BF01397280, w31} derivation of the uncertainty principle in quantum mechanics.

As it is well known, according to Heisenberg's uncertainty principle \cite{doi.org/10.1007/BF01397280} 
the product of the standard deviations  of position ($\sigma_x$) and momentum ($\sigma_p$) cannot be less than a non-zero constant   $\sigma_{x}\sigma_{p}\geq \frac{h}{4\pi}$, involving Planck constant $h$. Similarly,  the basic Gabor Uncertainty Principle  \cite[see][]{Gab46} states  that the product of
the uncertainties in frequency ($\sigma_f$) and time ($\sigma_t$) must exceed a fixed constant $\sigma_{f}\sigma_{t} \geq \frac{1}{4\pi}$.
As a direct consequence of this, it is impossible to know  the exact time and frequency of a signal simultaneously; hence, it is impossible to describe a signal as a point on the time-frequency (TF) plane.
TF analysis of time series is traditionally carried out by employing the Fourier spectra on successive sliding time windows  \cite[e.g., see][and references therein]{doi.org/10.1038/s41467-020-20539-9}.
Wide windows offer high frequency resolution but low temporal resolution, and vice versa; this is where the Heisenberg–Gabor uncertainty principle starts to have an influence.
The characterization of a time series in the frequency domain by means of the spectral density function $S(f)$, which establishes the \textit{distribution of the time series variance} at specific frequencies $f$, is one of the most used diagnostic tools for the identification of quasiperiodic fluctuations in a time series across disciplines. 
 The simplest estimator of the $S(f)$ is the periodogram  defined as the product of the time series sampling rate divided by the number of points and the square modulus of the discrete Fourier transform.
The major issues of the periodogram are well investigated \citep[see][and references therein]{doi.org/10.1029/2020JA028748}: (i)  Because of the finite frequency resolution, power leakage into adjacent bins occurs (ii) a bias in the estimate that was not known a priori, and which was dependent on the time series itself and (iii)the associated variance, that is equal to the estimate itself.

In order to get around above mentioned problems, multiscale approaches, which are also known as multiresolution methods, have been developed. An example of one of these methods is the continuous-wavelet transform \citep[CWT,][see]{daub92}. The CWT provides good relative temporal location of the signal, since it may either tighten or inflate a mother wavelet depending on the frequency of the signal \cite{Mal08}.

The CWT is capable of pinpointing the location of the oscillation in time, but as the frequency increases, it sacrifices its frequency resolution \cite{Tor98}.
However, in many instances, it is not possible to find difference between frequency components that are immediately adjacent to one another. Because of this, analyses are frequently carried out making use of a dyadic representation. One example of this is the dyadic discrete wavelet transform (DWT), in which the $T /2j$ wavelet amplitude coefficients arising from the $j$-th stage of the DWT, and $T = 2^{n}$ \cite{doi.org/10.1016/C2019-0-03019-1} are employed. On the other hand, this representation does not do a very good job of resolving the high frequencies.

 Our hybrid method, which is based on two-dimensional (2D) correlation analysis, was created in order to solve the problems that have previously been encountered \citep[see][]{10.1093/mnras/stx3137, 10.1515/astro-2020-0007}. It has been validated by the utilization of optical and photometric data obtained from several studies of active galactic nuclei, which are fueled by supermassive black holes \citep[see e.g.,][]{10.1093/mnras/stx3137,  10.3847/1538-4357/aaf731}. The fundamental feature of these data is that they have irregular sampling, with huge gaps, and signal with low amplitude, if it exists at all, is buried in the red noise.
 
 The 2D Hybrid method is able to utilize various  wavelets (e.g., CWT, DWT, Weighted wavelet transform, high resolution Superlets) in order to localize oscillations in period-period plane of time series in question.
The method  produces a contour
map of correlation intensity on a period-period plane defined by two independent period
axes corresponding to the two time series (or one). The map is symmetric and able to be integrated along any of the axes, resulting in a depiction of the level of correlation among oscillations that is similar to a periodogram. Therefore our approach could be interpreted as \textit{the two-dimensional distributions of correlation of the variance of time series in the time domain, which could also be projected into the one-dimensional domain}.  Our goal is to further illustrate the performance of the 2D hybrid technique and its application on time series of highly sampled very low frequency (VLF) radio signal amplitude data. 
%{  Our goal is to further illustrate the performance of the 2D hybrid technique and its application on time series of highly sampled very low frequency (VLF) radio signal amplitude data. These signals are used for ground-based observation of the Earth's lower ionosphere. They are emitted by worldwide located transmitters and propagate in the so-cold Earth-ionosphere waveguide, while numerous receivers record signal amplitude and phase at their locations. Variations in the characteristics of the recorded signal enable indirect detection of numerous phenomena, among which very significant are those related to natural disasters. Among others, a large number of studies based on the analysis of VLF signals investigate the possibility of the existence of earthquake precursors in the form of ionospheric perturbations. Those precursors are primarily associated with changes in the VLF signal amplitude and/or phase  \citep{bia06,bia01a,bia01b,roz04,zha20} and  the amplitude minimum time shift during solar terminator periods (the so-called “terminator time”) \citep{hay96,mau16,mol98,yam07,yos08}. In addition, the most recent research suggests that there are reductions in the noise of the VLF signal amplitude and phase a few tens of minutes before the observed type of disaster \citep{nin20a,nin21b,nin22b}.
In this study, we present computations of 2D correlation maps of  the VLF signal amplitude oscillations before, during and after the earthquake nearby city Kraljevo in Serbia  on November 3, 2010.
As perturbation of  VLF signal amplitude associated with an occurrence of earthquakes, this application presents an opportunity for the acquisition of novel insights.

%%%%%%%%%%%%%%%%%%%%%%%%%%%%%%%%%%%%%%%%%%
\section{Materials }

{
This study is based on the processing of data recorded in the low ionosphere monitoring by the 20.27 kHz VLF radio signal emitted by the VLF transmitter {\citep[which name is ICV, see e.g.,][]{zhao19}} located in Italy (Isola di Tavolara, Sardinia, latitude 43.74 N, longitude 20.69 E) and recorded by the Absolute Phase and Amplitude Logger (AbsPAL) receiver in Belgrade, Serbia (latitude 44.8 N, longitude 20.4 E). The time intervals are determined in relation to the time of the earthquake that occurred near Kraljevo (Serbia) on November 3, 2010 at 00:56:54.4 Universal Time (UT). In Figure \ref{map}, we show the path of the observed VLF signal and the location of the Kraljevo earthquake epicentre \cite{kne13} for which the data are given in \url{http://www.emsc-csem.org/Earthquake/}.

\begin{figure}[H]
%\begin{adjustwidth}{-\extralength}{0cm}
\includegraphics[width=0.7\linewidth]{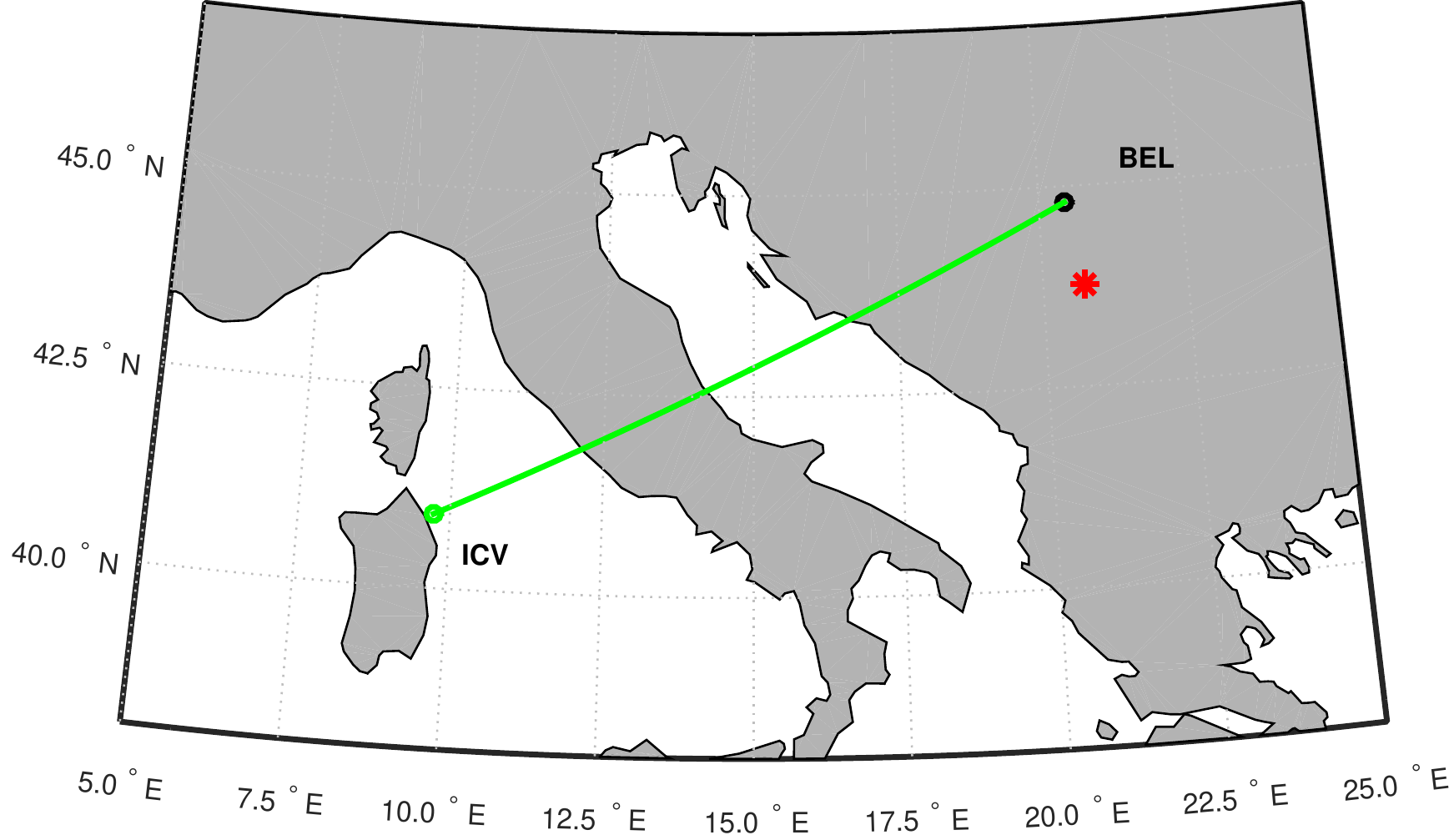}
%\end{adjustwidth}
\caption{Propagation path of the VLF signal recorded by the Belgrade receiver station (BEL) in Serbia and emitted by the  ICV transmitter in {Isola di Tavolara, Sardinia,} Italy (solid line). Location of Kraljevo earthquake epicentre is shown as star. \label{map}}
\end{figure}  

In this study, we consider data series showing VLF signal amplitude values recorded with a sampling of 0.1 s in five time intervals starting 2 h and 1 h before the Kraljevo earthquake, at the moment of its occurrence, and 1 h and 2 h after that time. In order to compare the obtained results with those relevant for periods without seismic activity, we additionally analyse the time intervals in the same season period but about one year earlier (1-2 November, 2009). These periods are chosen to exclude the effects of daily and seasonal changes that are visible in the VLF signal amplitudes and that may affect the observed comparison. In addition, it was taken into account to eliminate the potential influences of other natural phenomena with origin in the atmosphere and from space as well as non-natural causes of variations in the emission and reception of the considered signal (these influences are described in detail in \cite{nin20a} and \cite{nin22b}). For this reason, the reference intervals are chosen in periods when no significant disturbances in meteorological, geomagnetic and space weather conditions were recorded, and when approximately the same values of the amplitude and its noise were recorded as in the quiet period before the Kraljevo earthquake.

}
\section{Methods}

Here, we are presenting a detailed view of 2D Hybrid method, where the main steps of this process are depicted, while we provide excerpts of the algorithm in pseudocode for revealing additional important details (see  Algorithm \ref{alg:cap}, in Appendix
).
{
To start, we review some of the fundamental ideas and information associated with wavelet analysis (see the Section \ref{sec:wave}). The overall concept of the 2DHybrid method is presented in Section \ref{sec:2D}, along with a detailed description of the period's uncertainty (Section \ref{sec:uncer}) and significance (Section \ref{sec:sign}) estimate. In conclusion, the Section \ref{sec:wave2} presents the wavelet that was implemented in our 2DHybrid method.}

\subsection{Summary of wavelet concept}\label{sec:wave}
A wavelet function (abbreviated wavelet) is a function belonging to the space of all square-integrable functions $\psi \in L_{2}(\mathbb{R})$, averaged $\int_{\mathbb{R}} \psi=0$, and normalized $\| \psi\|=1$ \cite[see more details in][]{daub92, Vidak99}.
When we compare the wavelet approach to the Fourier method, we find that the Fourier analysis disassembles a signal into a set of sinusoids defined with distinct frequencies, but the wavelet method unwinds a signal into the shifted or scaled shapes that originate from a mother wavelet.
 Wavelet maps a signal into a time-scale plane which  is the same as the time–frequency plane in the short-time Fourier transform, so that each scale   represents a certain frequency range of the time–frequency plane \cite{10.3390/s19040935}. 
Given the signal $f(t)$, its CWT at time $u$ and scale $s$ is defined as:

\begin{linenomath}
\begin{equation}
CWf(u,s)=\frac{1}{\sqrt{s}}\int^{+\infty}_{-\infty} f(t)\psi^{\star}\left(\frac{t-u}{s} \right) dt\,\,\,\,
s\in\mathbb{R}-{0}, u\in \mathbb{R}
\end{equation}
\end{linenomath}

 The result of the CWT is a matrix (scalogram) filled with wavelet coefficients located by scale and position:

 \begin{linenomath}
\begin{equation}
\mathcal{S}(s)=\|CWf(u,s)\|=\Bigg(\int^{+\infty}_{-\infty} \|CWf(u,s)\|^{2} du\Bigg)^{1/2}.
\end{equation}
\end{linenomath}

\noindent The aforementioned equation could be interpreted as the amount of energy, denoted by $CWf$ present at scale $s$, provided that the condition $\mathcal{S}(s) \geq 0$ holds. When seen in this context, the same equation enables us to determine the scales that contribute the most to the overall energy of the signal.

In most cases, we are only interested in searching for oscillations inside a predetermined time span such as $[t_{min}, t_{max}]$, and because of this, we can define the windowed scalogram that corresponds to this interval as follows:

\begin{linenomath}
\begin{equation}
\mathcal{S}_{[t_{min}, t_{max}]}{(s)}=\left.\|CWf(u,s)\|\right\vert^{t_{max}}_{t_{min}}=\Bigg(\int^{t_{max}}_{t_{min}} \|CWf(u,s)\|^{2} du\Bigg)^{1/2}
\end{equation}
\end{linenomath}

In actuality, any time series $f$ that is appropriate for wavelet analysis needs to be defined over a limited time interval $[t_{min}, t_{max}]$  and sampled with a specified resolution in order to obtain discrete data.
In terms of sampling, any discrete signal can be studied in a discrete domain by utilizing discrete wavelets, or in a continuous manner by employing neural and Gaussian process models of discrete series with gaps.

If we are given two time series, $f$ and $f^{\prime}$, we can examine their respective scalograms, $\mathcal{S}$ and $\mathcal{S}^{\prime}$, to determine whether or not they exhibit patterns that are comparable to one another. For instance, we are able to perform an absolute comparison of scalograms using the formula  $\| \mathcal{S}-\mathcal{S}^{\prime}\|$ \citep{Bol14}. 

\subsection{General description of 2DHybrid method}\label{sec:2D}
{Now, we are ready to present our 2D Hybrid approach,} which  compares the scalograms of two different (or one) series by using correlation. Given a scalogram $\mathcal{S}$ with dimensions  $M\times N$ and another scalogram $\mathcal{S}^{\prime}$ with dimensions  $P\times Q$,  the two-dimensional cross-correlation ($\star$) of these scalograms is the matrix $\mathcal{C}=\mathcal{S}\star\mathcal{S}^{\prime}$ with dimensions $ [M+P-1]\times [N+Q-1]$,  which has following elements:

\begin{linenomath}
\begin{equation}
\mathcal{C}_{(k,l)}=\sum^{M-1}_{m=0}\sum^{N-1}_{n=0}
\mathcal{S}(m,n)\overline{\mathcal{S}}^{\prime}(m+k,n+l)
%Tr\Bigg(\tilde{\mathcal{S}}\tilde{\mathcal{S}^{\prime}_{kl}}^{\dagger}\Bigg)\,\, k={1,...,M+P-1}, l={1,...,N+Q-1}
\end{equation}
\end{linenomath}

\noindent where  $\overline{\mathcal{S}}^{\prime}$ stands for complex conjugate of ${\mathcal{S}}^{\prime}$ and $-(P-1) \leq k\leq M-1, -(Q-1)\leq l\leq N-1$.

As the cross correlation of scalograms is defined on the field of complex numbers, both the real and imaginary components of the complex cross correlation function are referred to as the synchronous and asynchronous 2D correlation spectra, respectively \cite[see][]{Nod15}.
Since we are only interested in physical phenomena whose correlation can be tracked in the field of real numbers, we only supply the mathematical formulation of synchronous 2D correlation spectra \cite{Sch04}

\begin{linenomath}
\begin{equation}
\mathcal{M}=\frac{Cov(\tilde{\mathcal{S}},\tilde{\mathcal{S}^{\prime}})}{\sigma \sigma^{\prime}}
\end{equation}
\end{linenomath}

\noindent where $Cov$ stands for covariance and $\sigma$ and $\sigma^{\prime}$ are standard deviations of scalograms  $\tilde{\mathcal{S}}$ and $\tilde{\mathcal{S}^{\prime}}$ of two time series, respectively.

Notably, in the discrete formulation of cross correlation, a synchronous 2D correlation map is simply defined as the inner product of the
$\tilde{\mathcal{S}}$ and $\tilde{\mathcal{S}^{\prime}}$ \cite{Sch04}:

\begin{linenomath}
\begin{equation}
\mathcal{M}=\tilde{\mathcal{S}}^{T} \tilde{\mathcal{S}^{\prime}}
\end{equation}
\end{linenomath}

The resemblance between  oscillations in two different (or one) time series is measured via a 2D correlation map {(which we will also refer to as a heatmap)}. A high positive correlation value shows that periodic signals vary in a coordinated manner, implying that the signals have a common or related origin \citep{10.1093/mnras/stx3137}.

The two-dimensional correlation map is presented as a contour map of correlation strength on a period plane that is defined by two axes that are independent of one another. When done in this fashion, plotting a synchronous spectrum results in a map that is symmetric in relation to the primary diagonal line of the map. The correlation of two (or one) time series at the same periods refers to the intensity of the correlation that may be found at the main diagonal of the map. As a result, peaks that are located on the main diagonal line are referred to as auto-peaks. The intensities of auto-peaks are representative of the total extent of the signals' dynamic fluctuations \citep{10.1093/mnras/stx3137}.
It is important to note that two-dimensional correlation map can be summed with the absolute value of $C(k,l)$ along any of the dimension. This is because negative correlation can also appear for some signals which should not be canceled in the summation.
{This integration}  provides an interpretation similar to a periodogram, with the horizontal axis counting periods and the vertical axis standing for the degree of correlation peaks.

\subsubsection{Concept of estimating uncertainity of detected periods}\label{sec:uncer}
 In order to find the uncertainty of detected periods,  we first calculate the full width half maximum of the peak in a periodogram-like image, and then  use the \texttt{mquantile} module in \texttt{Python} to estimate points that fall between the 25th and 75th quantiles. The upper and lower error estimates are represented by these points. {The reason behind using quantiles is that peaks in periodogram -like structure do not conform to the theoretical normal distribution, actually they are skewed.  In
this particular situation, quantiles provide more
suitable information than standard deviation. The sample quantile is based
on order statistics and calculated regardless of
underlying distribution. The $p$-th quantile of a set of values
represents a summarizing quantity having less than or
equal to p, where, $0\leq p \leq 1$.
  Similarly to median which is the value bellow which 50$\%$ of all values in the sample lie, we might define that the first quartile (25th quantile)  as the value below which 25$\%$ of all values in the sample lie and the third quartile (75th quantile) as the value below which 75$\%$ lie.}

 Our period error method is inspired on 'post mortem analysis' by \cite{Schw91}, which requires the so called Mean Noise Power Level (MNPL) in the vicinity of detected period. The 1-sigma confidence interval on period then is equal to the width of the line at the period – MNPL level.

 \subsubsection{Concept of estimating significance of detected periods}\label{sec:sign}
The significance of detected period $\sigma_P$, we  estimated following the approach outlined in \cite{John19}.
After shuffling the dates of each and every observation and its magnitude, the period was recalculated across this newly updated data set, and the power of the highest peak in this uncorrelated data set was compared to that of the initial simulated data. After performing this procedure a total of one hundred times (presumably due to large computing time needed for highly sampled VLF signals), the significance level was finally calculated as:
\begin{equation}
\sigma_{P}=\frac{x}{N}
\end{equation}
where $x$ represents the number of times that the peak power of the period in the original data was greater than that of the uncorrelated ensemble and $N$ is the total number of shuffles (100). This formula therefore has a maximum of 1, corresponding to a 100 percent recovery rate.
{When multiple periods are found in an original curve, the significance of each peak is measured by comparing the power of peaks in shuffled curves found at the place of detected periods to the power of peaks in the original curve.}
 \subsubsection{Wavelets used in 2DHybrid method}\label{sec:wave2}

The effective implementation of the 2D Hybrid approach is going to be demonstrated in the next Section by making use of Weighted Wavelet Z-transform (WWZ) wavelets \cite{Fost96}. This wavelet approach can be utilized on data that has been sampled both regularly and irregularly.
The WWZ wavelets are defined on a basis that consists of {functions}:  $\cos[\omega(t-\tau)], \sin[\omega(t-\tau)]$, $I(t)=1$. In addition, the projection of data via WWZ makes use of weights that take the form $exp(-c\omega^{2}(t-\tau)^{2})$, where $c$ is a parameter that can be adjusted according to the data set.

The tuning constant $c$, whose value determines the window's width, can have a variety of  choices. For instance, the value $c = 0.0125$ was initially suggested in \cite{Fost96}, with the intention of improving the time resolution on shorter parts of the data.
However, the value of $c$ might go as high as 0.005, and it is used to  in order to improve frequency resolution.

In the former scenario, this translates to the wavelet  decaying by $e^{-1}$ in $\sim$1.4 cycles, while in the later scenario, this translates to the wavelet deteriorating in  $\sim$2.4  cycles.
In our study we used $c\sim 0.003$. This value can be compared to e.g. $c = 0.001$ used in \cite{Temple05} and  \cite{Y13} for longer data sets than the one used here.

Because each of the analyzed time series has 36 000 points, the application of our 2D Hybird method takes 45 minutes to complete on the \texttt{google.colab} computing platform, which has a CPU: 1$\times$ single core hyper threaded Xeon Processors 2.2 GHz (1 core, 2 threads) and RAM: $\sim$ 13 GB. It takes about 70 hours on the same platform to calculate the significance of a detected period using 100 artificial time series.

\section{Results and Discussion}

\begin{figure}[H]
\begin{adjustwidth}{-\extralength}{0cm}
\includegraphics[width=6cm]{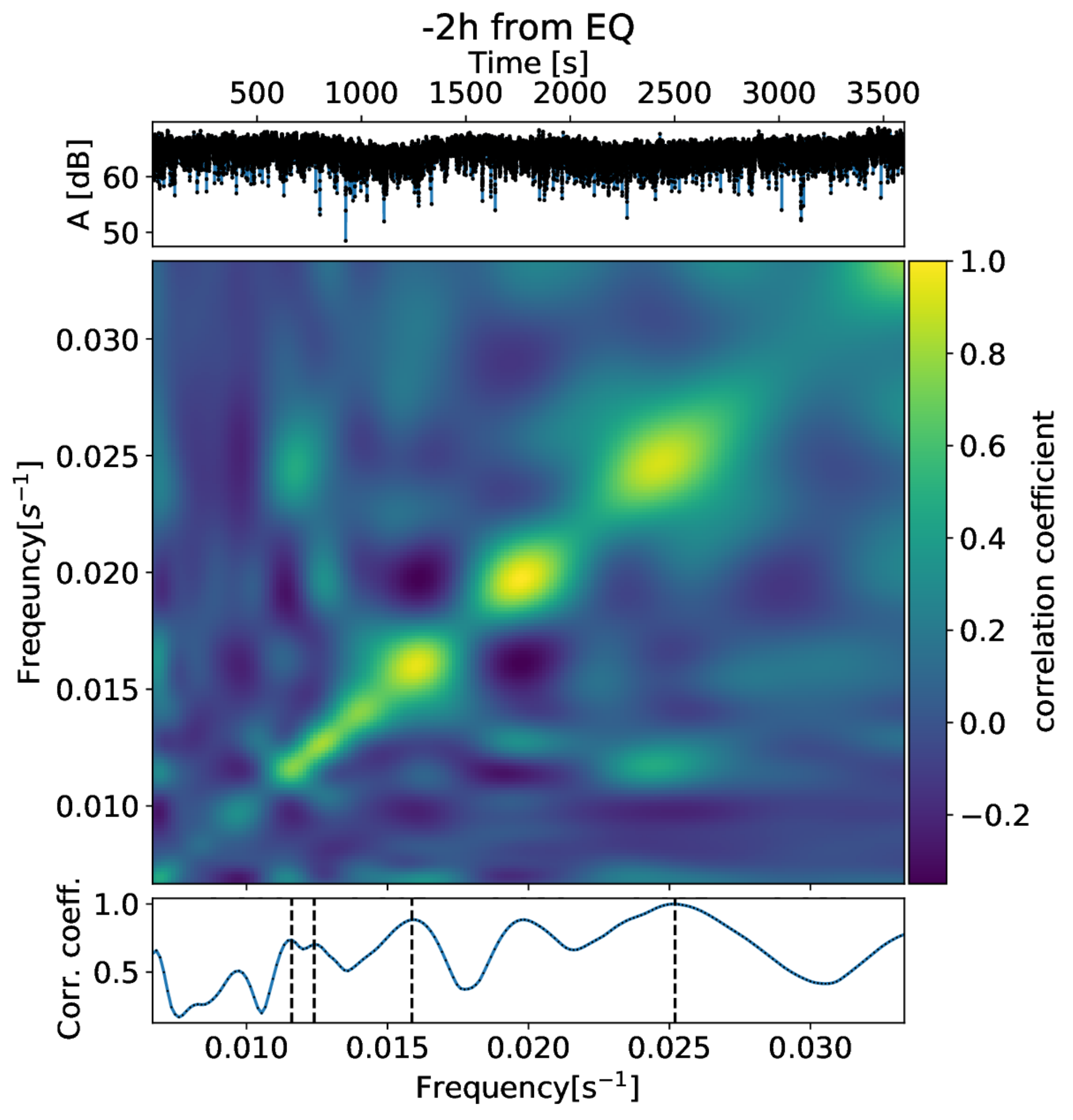}
\includegraphics[width=6cm]{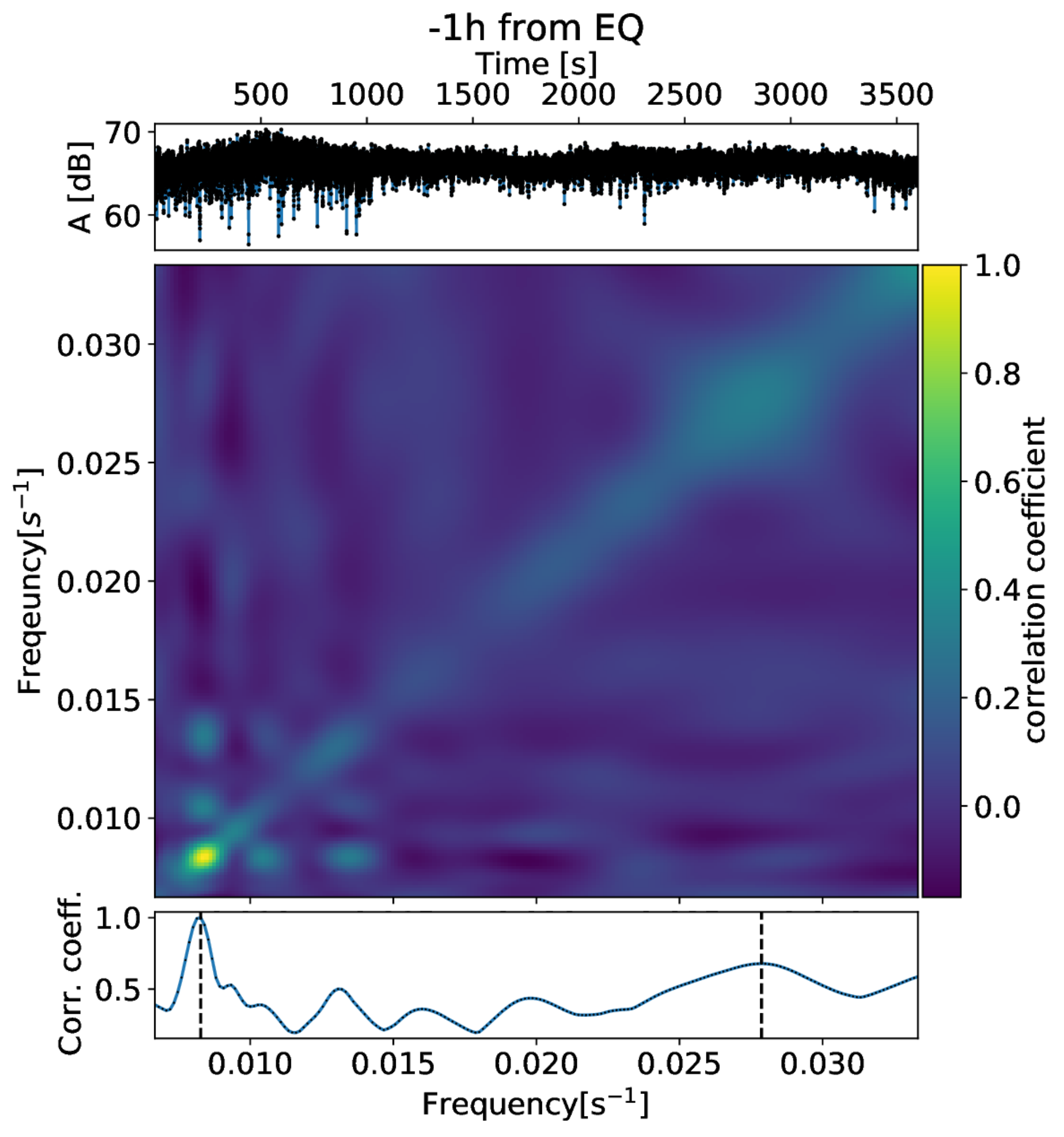}\\
\includegraphics[width=6cm]{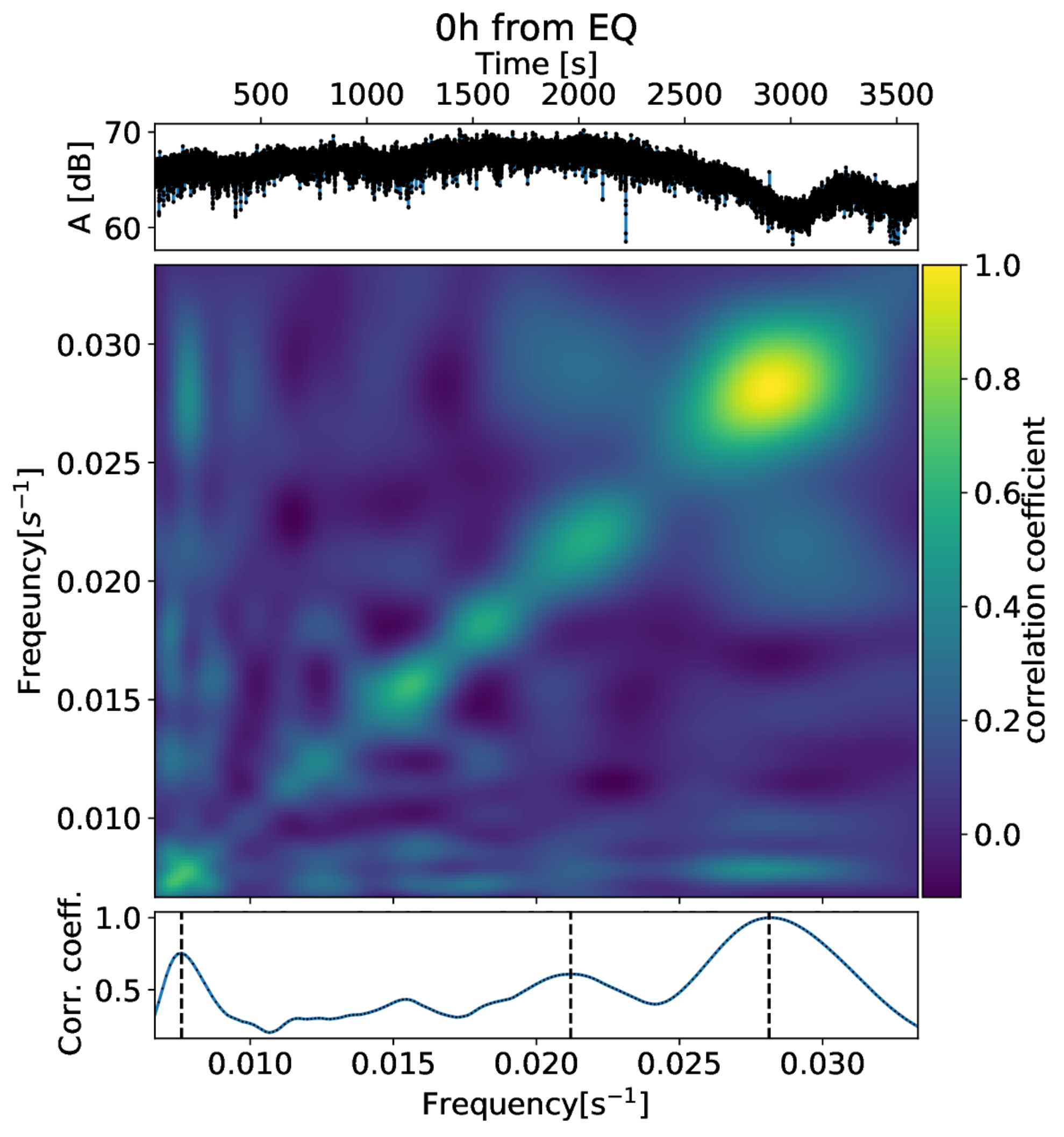}\\

\includegraphics[width=6cm]{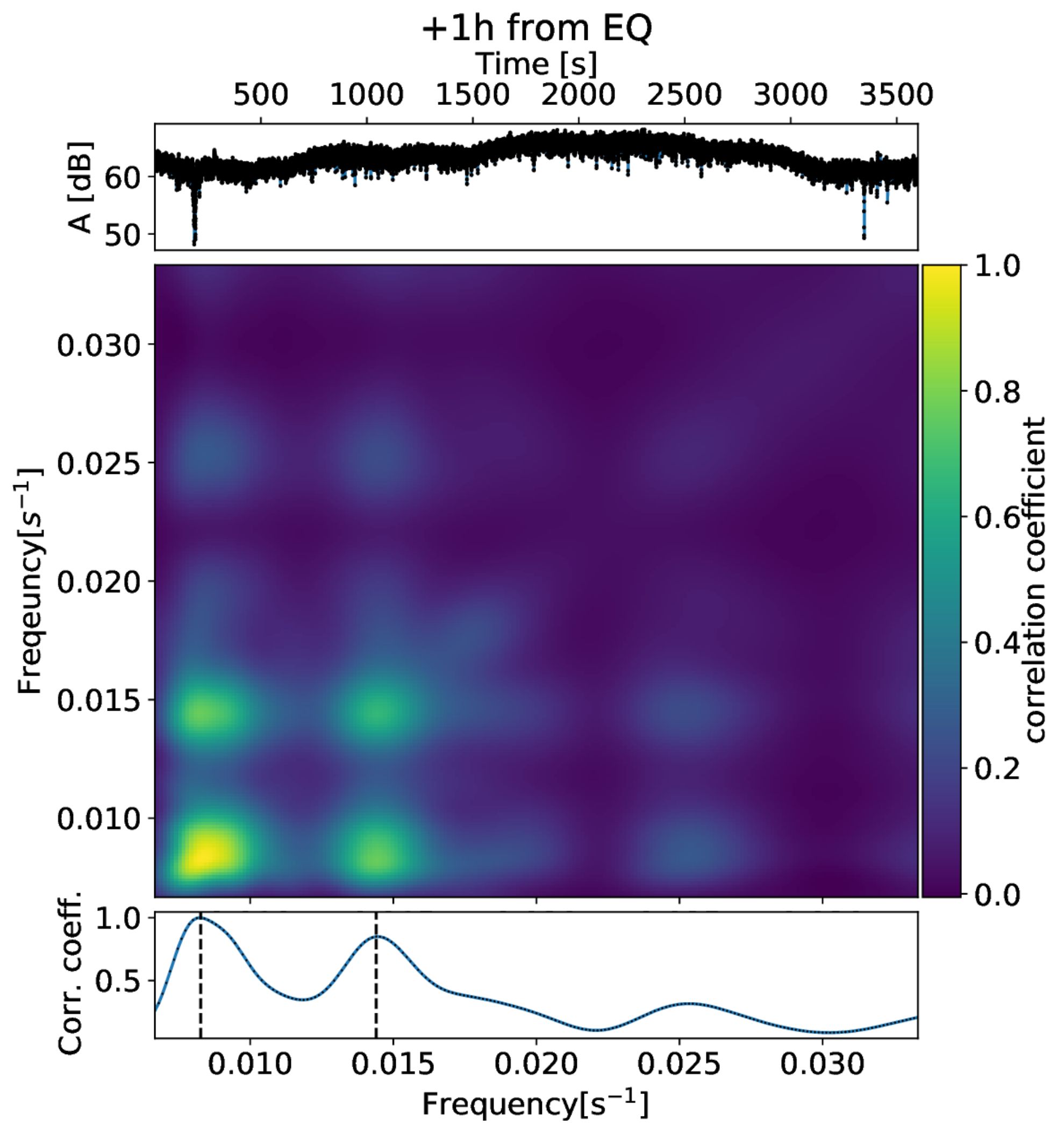}
\includegraphics[width=6cm]{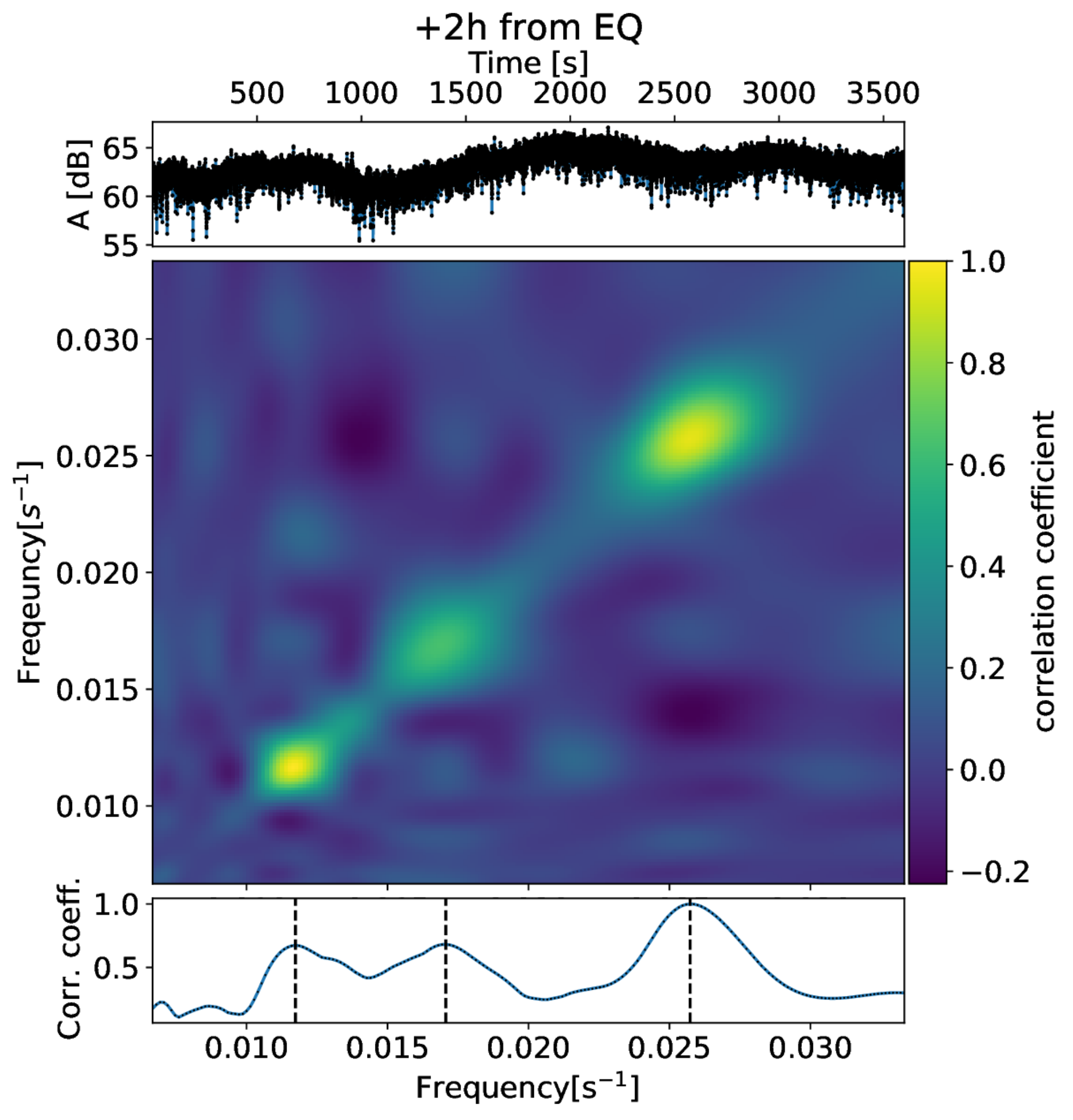}
\end{adjustwidth}
\caption{2D Hybrid maps of oscillations in VLF signal amplitude data. Each plot shows portion of time series relative to beginning of earthquake (top marginal panel), 2D Hybrid map (middle panel) and projected 2D Hybrid map (bottom marginal plot), colorbar represents correlation coefficients on map. Top row: time series 2 h before earthquake (left) and 1 h before earthquake (right); middle row: 0 h from earthquake; bottom row: 1 h after earthquake (left) and 2 h after earthquake (right). \label{figg}}
\end{figure}

\begin{figure}[H]
\begin{adjustwidth}{-\extralength}{0cm}
\includegraphics[width=6cm]{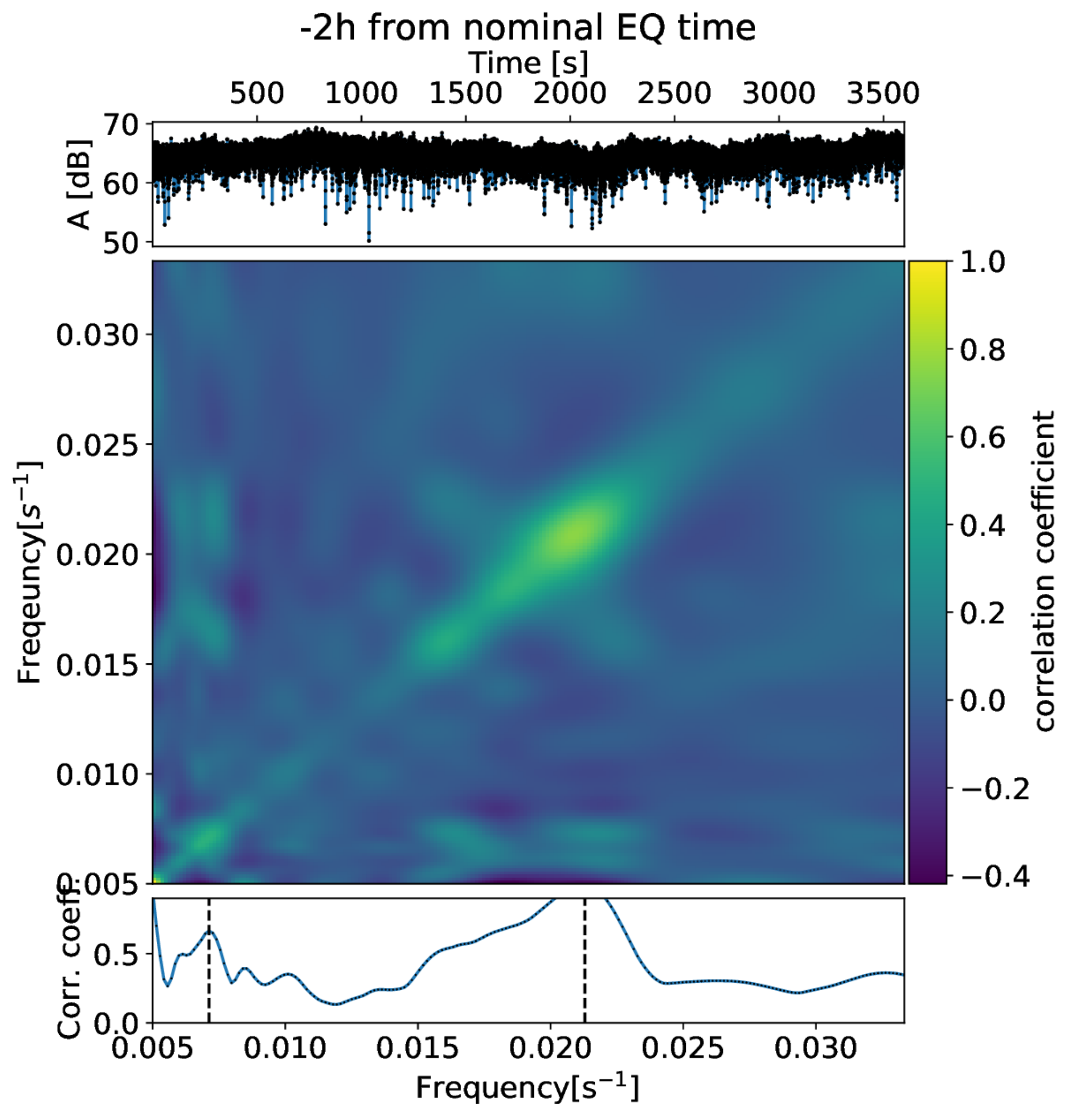}
\includegraphics[width=6cm]{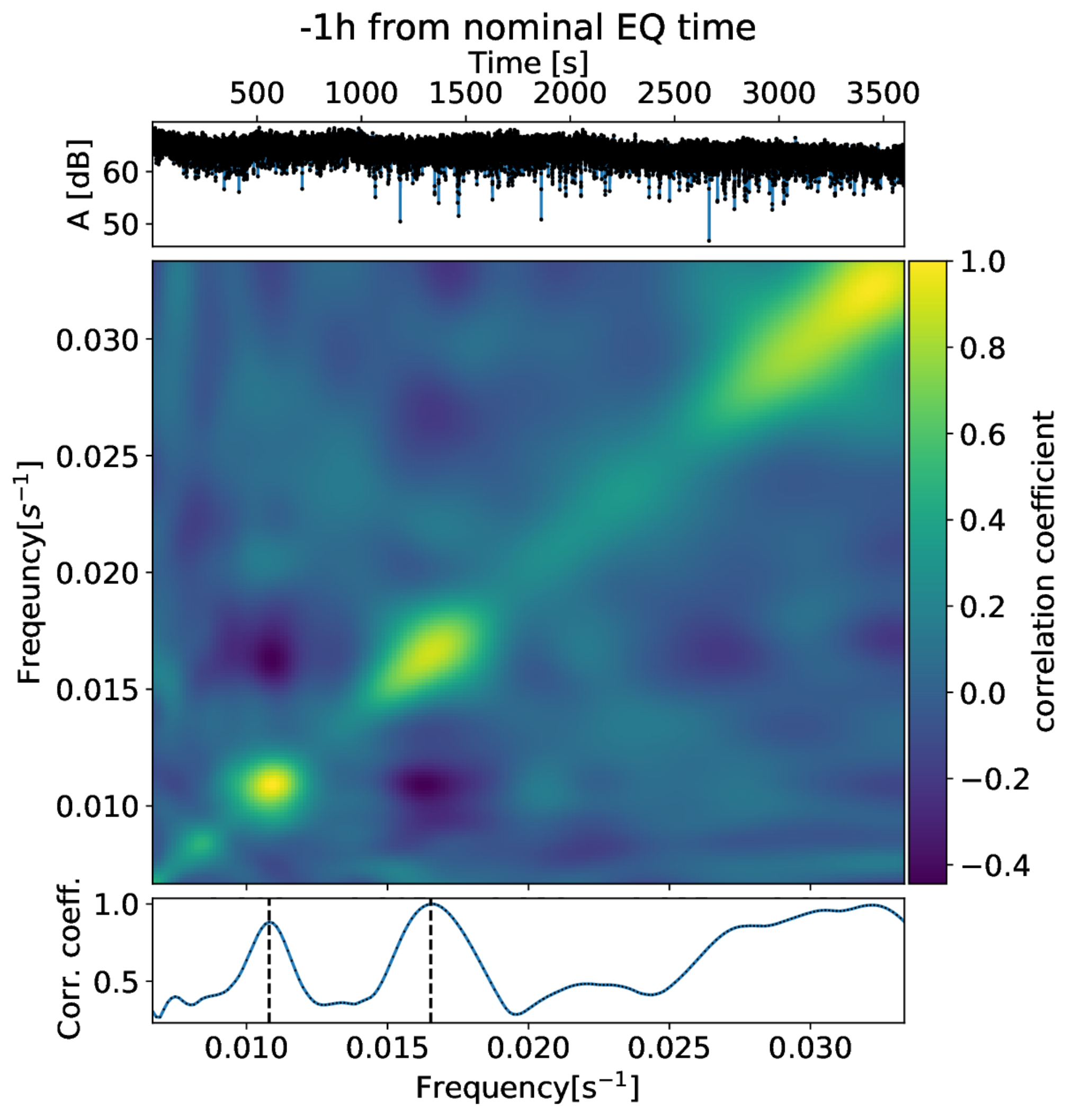}\\
\includegraphics[width=6cm]{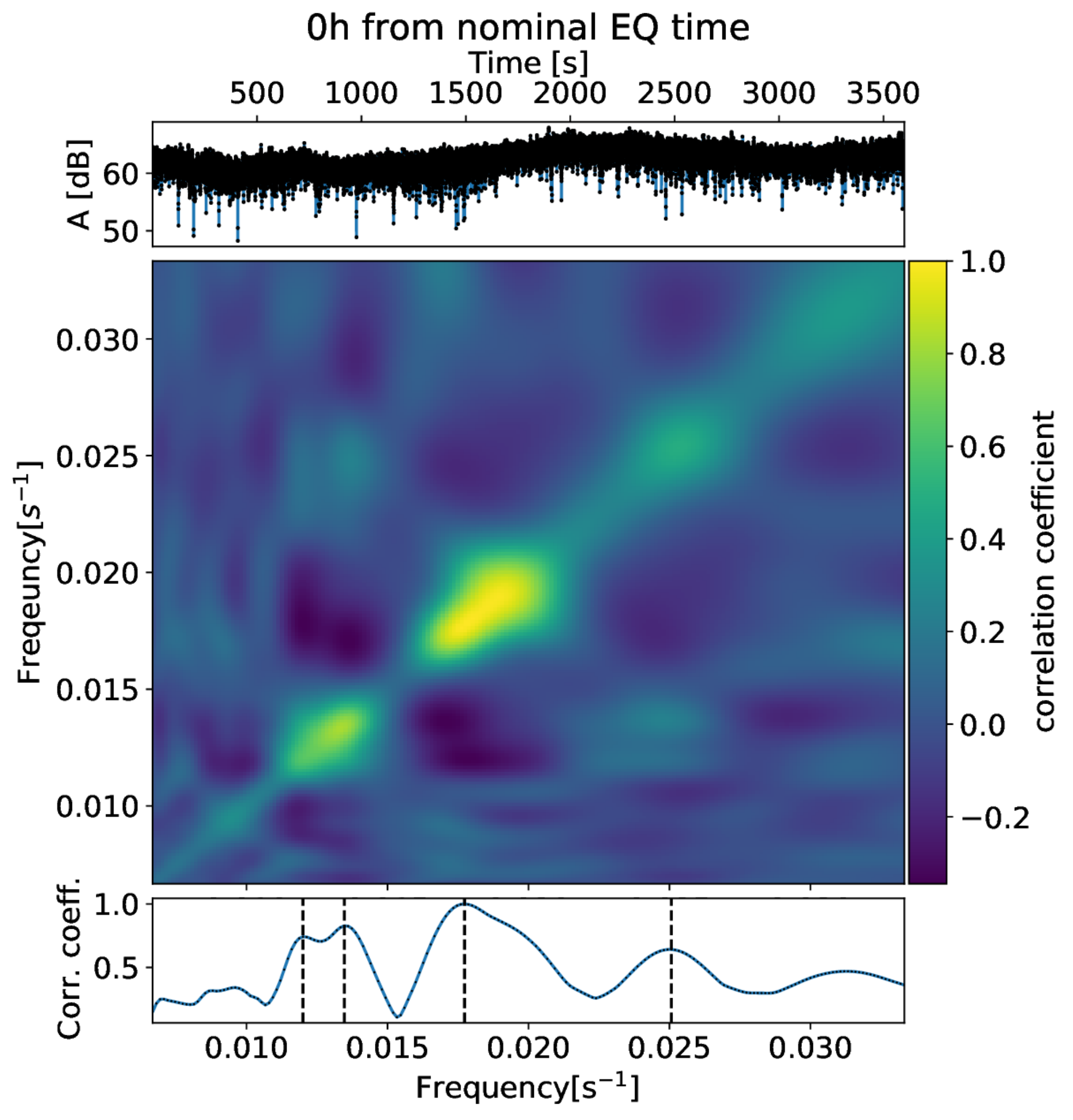}\\

\includegraphics[width=6cm]{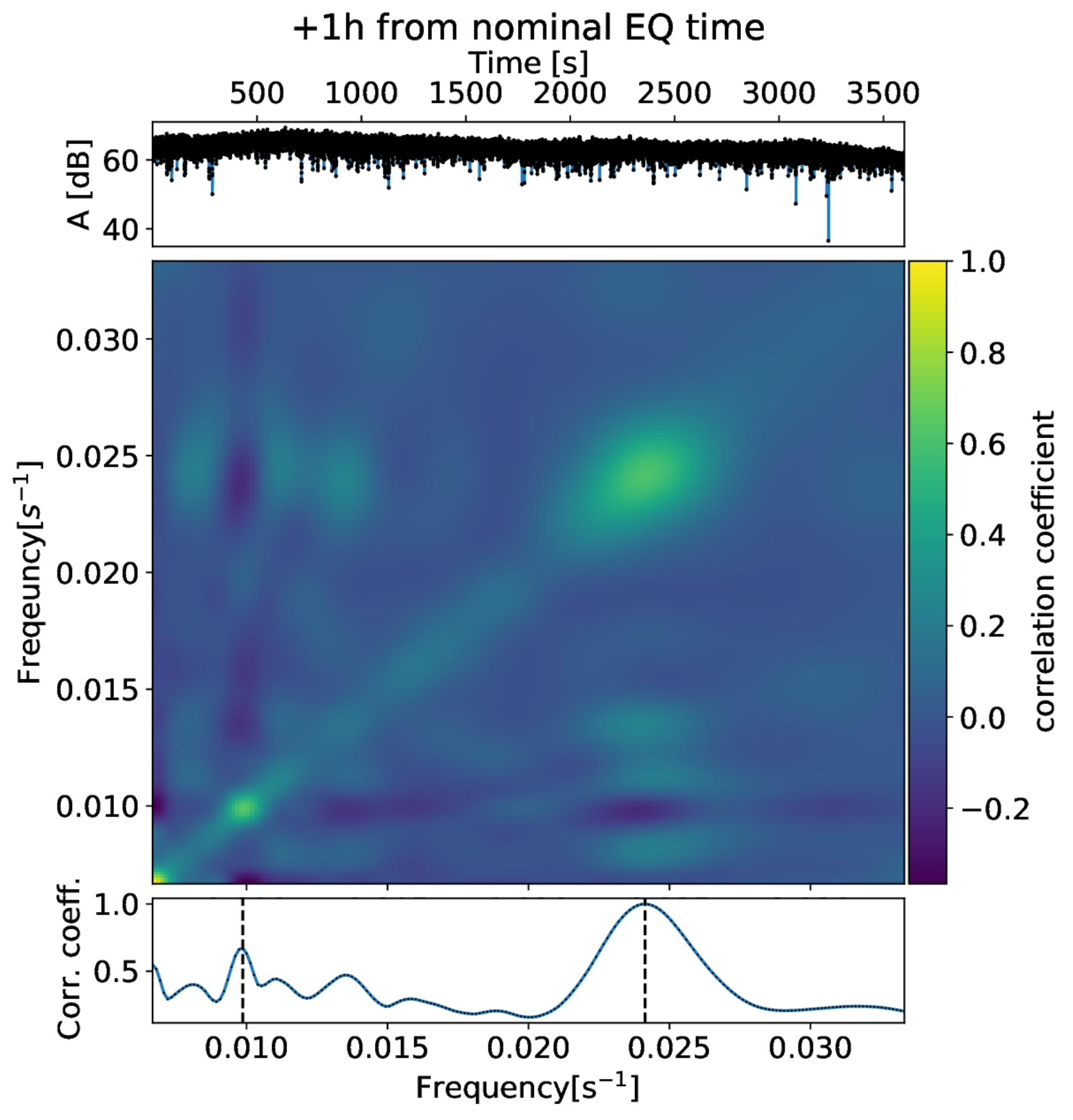}
\includegraphics[width=6cm]{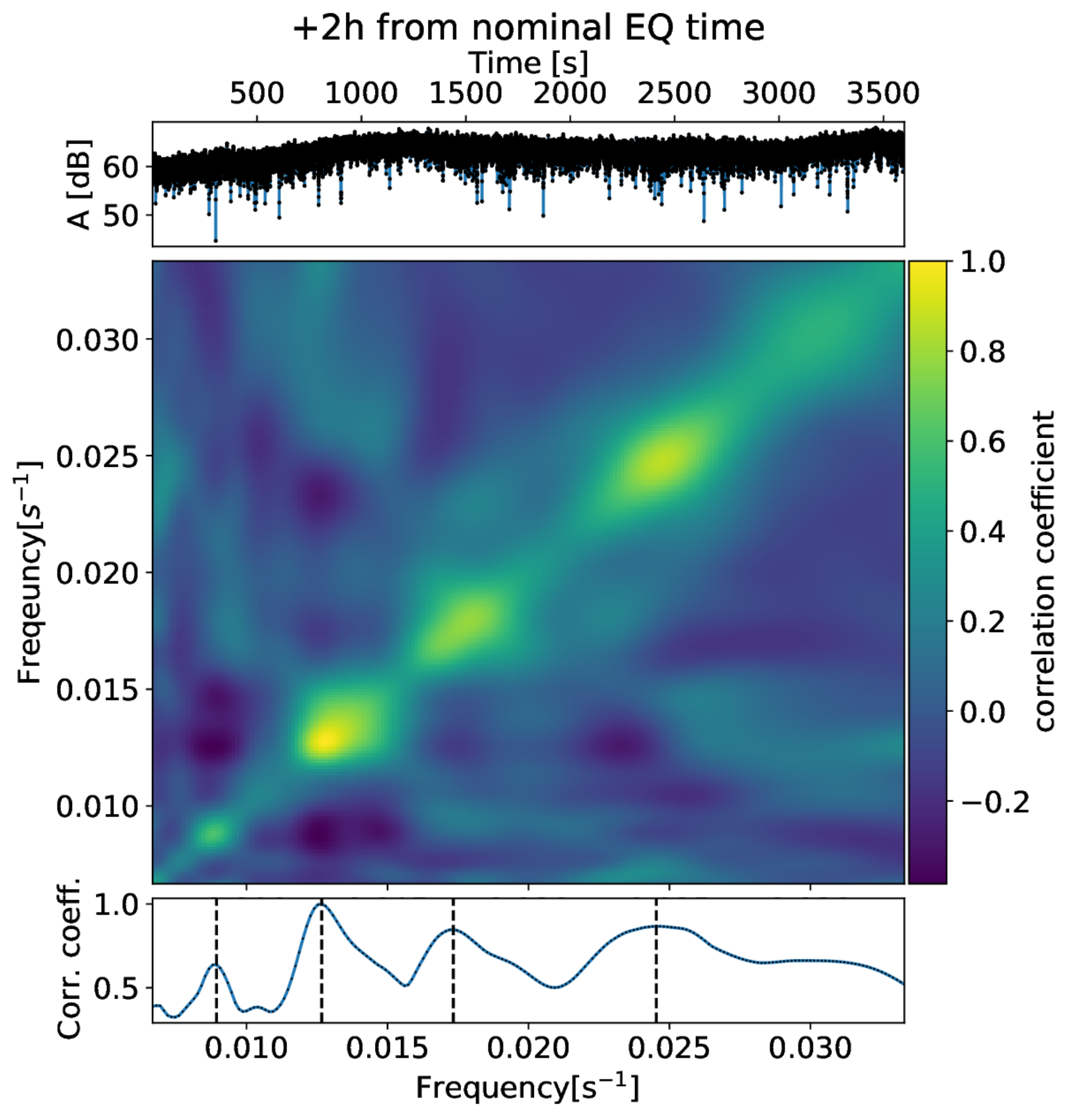}
\end{adjustwidth}
\caption{ The same as Figure \ref{figg} but for date 1 year earlier.\label{figg2}}
\end{figure}

As the amplitude fluctuation of the VLF signal during the Kraljevo earthquake is known to be difficult to analyze using Fourier periodicity, we decided to demonstrate our method using this data.
We computed 2D hybrid maps and their integrated versions for each segment of the time series for the date of the earthquake occurrence (see Figure \ref{figg}) and for the same date but one year earlier as a control case (see Figure \ref{figg2}).
 For each time series segment we kept the same parameters of WWZ wavelets: $c=0.003$, range of frequencies [1/150,1/30 ]\,$s^{-1} $, and number of points in frequency grid (200). 
 Table \ref{tabb} provides a summary of the detected periods for the date of earthquake occurrence, whereas Table \ref{tabb2} provides the values for those detected periods for the control date one year earlier.
 \begin{table}[H]
\caption{Summary of detected periods on the date of earthquake. Periods are measured from integrated version of 2D Hybrid map  (bottom marginal panels) in Figure \ref{figg}. Columns: series part with respect to earthquake beginning time,  detected periods, lower and upper errors.\label{tabb}}
	\begin{adjustwidth}{-\extralength}{0cm}
		\newcolumntype{C}{>{\centering\arraybackslash}X}
		\begin{tabularx}{\fulllength}{CCCCC}
			\toprule
			{Series ID}	& {Period [s]}	& {-err [s]}     & {+err [s]}
			 & {significance [$\%$]}
			\\
			\midrule
			-2 h 
			&147.06  &   56.8&      60.0 &99\\
			& 86.2   &    4.2   &    5.8& 99\\
				&80.6   &    15.5   &   18.8&99\\
			&	63.02 &     11.3  &    12.8 &99\\
			&50.33 &      9.7      &10.7& 99\\
			\hline
				-1 h & 121.0 &       10.0  &     0.9&\\
&35.9 &      0.7    &   0.2&99\\
\hline
				0 h &	131.6       &7.4 &1.1&\\
&47.2&        0.9 &   0.3&99\\
&35.5 &        0.5 &   0.2&99\\
\hline

+1 h &121.0 &      7.8 &        1.3 &99\\
&69.4  &      2.5 &        1.7&\\
\hline
+2 h& 85.2 &      3.9&      1.0&99\\
 &58.6 &       2.0 &      0.25&99\\
 &38.9 &       0.8 &      0.2&99\\
			\bottomrule
		\end{tabularx}
	\end{adjustwidth}
%	\noindent{\footnotesize{\textsuperscript{1} This is a %table footnote.}}
\end{table}

 The heatmaps  show the prominent oscillations in time series segments which are plotted on the top of heatmap.
 The frequencies are displayed in $\mathrm{s}^{-1}$ along both the x and y axes of the plot. The degree of correlation between the oscillations in the time series is represented by the color of the heatmap cell for each pair of values (x, y). According to the colorbar scale that can be seen to the right of each plot, the hues of the heatmap are related with the correlation coefficients.
It is important to note that the topology of heatmaps varies across different time series segments.
 
\begin{table}[H]
\caption{Summary of detected periods in control case corresponding to the same date as earthquake occurence but one year earlier. Periods are measured from integrated versions of  2D Hybrid maps (bottom marginal panels) in Figure \ref{figg2}. Columns: series part with respect to nominal earthquake beginning time,  detected periods, lower and upper errors.\label{tabb2}}
	\begin{adjustwidth}{-\extralength}{0cm}
		\newcolumntype{C}{>{\centering\arraybackslash}X}
		\begin{tabularx}{\fulllength}{CCCCC}
			\toprule
			{Series ID}	& {Period [s]}	& {-err[s]}     & {+err [s]}
			 & {significance [$\%$]}
			\\
			\midrule
			-2h& 140.4& 2.7& 2.7&99\\
&47 &0.07&0.07&99\\
			\hline
				-1 h 	&92.6& 3.6&1.1&99\\
&60.5& 1.5& 0.5&99\\
\hline
				0 h  &83.3& 2.9& 0.9&99\\
&74.2& 2.3& 0.7&99\\
&56.4& 1.3& 0.4&99\\
&39.9& 0.8& 0.03&99\\
\hline

+1h &101.3& 5.2& 0.3&99\\
&41.4& 0.7&0.2&99\\
\hline
+2h& 111.9& 6.5& 0.1&99\\
&79& 2.6& 0.3&99\\
&57.7& 1.5& 0.2&99\\
&40.8& 0.7 &0.2&99\\
&33.8& 0.5&0.2&99\\
\hline

			\bottomrule
		\end{tabularx}
	\end{adjustwidth}
%	\noindent{\footnotesize{\textsuperscript{1} This is a %table footnote.}}
\end{table}

%%%%%%%%%%%%%%%%%%%%%%%%%%%%%%%%%%%%%%%%%%

At 2 hours before the earthquake (shown in Figure \ref{figg} top row, left plot) we observe several signals of $\geq 80 $\,s with correlation coefficients larger than 0.5 (shown in the bottom marginal plot of heatmap integrated projection), but there is only one peak  above 100\,s (located at 147 s, see also Table \ref{tabb}).
All of the other oscillations have been muted by the passage of time, so that only the oscillation of  120\, s that occurred 1 h before the earthquake is clearly visible in the top right panel of  Figure \ref{figg}. 

During the earthquake, at 0 h from the event (middle plot), our approach records an oscillation at 131 seconds, with an amplitude that is approximately 20 percent lower than the peak that was recorded at 120 seconds during the hour-long interval preceding the earthquake (top right plot). In addition, there is one more signal at 35 s (middle plot).

After an hour has passed since the occurrence (bottom left plot), the oscillation at 121 s and 70 s once again appears.
Finally, two hours after the earthquake (bottom right plot), the signals lasting longer than one hundred seconds vanished, and the graph once again began to show oscillations in the 80-90 s range, just as it had done in the period lasting two hours before the earthquake (top left plot).

One hour after the event, once again appears oscillation at 121 s, and 70 s.
Finally, two hours after the earthquake the signals above 100 s disappeared, and once again comes to display oscillations around $\sim 80-90$ s as in the period of two hours before earthquake.
Interestingly, only  time series segments corresponding to -2 h (the top left panel) and +2 h (bottom right panel) have a topology that is comparable to one another.

{As it can be seen in Table \ref{tabb}, there are some ionospheric periodical oscillations of the order of one to two minutes. It is a question of this oscillation and its physical origine. One of the possibilities is that the electron density in the ionosphere is following periodical changes in the electromagnetic field which are registered close to the epicenters of earthquakes (see e.g.
\cite{shi07,ots06,khe09,hek21}), or that be generated by acoustic waves (see e.g. simulations in \cite{shi07}). However, the true nature and physical background of the short-period oscillations given in Table 1 should be investigated in more detail, which is out of the scope of this paper.}

In the scenario when there is no earthquake (the same date but one year earlier), the topology of the 2D hybrid maps Fig \ref{figg2} looks very different from the topology of the maps corresponding to the record for earthquake date (Figure \ref{figg}). The primary diagonal is where the majority of the correlation clusters are arranged. The  VLF signal amplitude variation can be seen as a switching between a dominated correlation cluster at -2 h (top row-left), +1 h (bottom row, left)  and a more granular structure (at +2 h (bottom row-right) and -1 h (top row-right). This variation on topology  can not be  seen during earthquake occurence (see Figure \ref{figg}). On top of this,  we observe a dominating core cluster is present at 0 h  (middle panel). On the other hand, the values of detected  periods are less than 111 seconds (at +2 h, +1 h, 0 h, and -1 h, see Table \ref{tabb2}). It is interesting to note that a period of 140 s is captured in the -2 h time series segment, which is comparable to a period of 147 s that was captured in the -2 h segment when the earthquake occurred (Table \ref{tabb}).

{A comparison of the obtained results with those given in \cite{nin20a} (based on the application of the Fast Fourier transform (FFT) to the data in the relevant time intervals of 1 h) shows that the agreement is better before the earthquake. For the first observed interval starting 2 h before the earthquake, the agreement is good for the obtained values below 1.5 min, while, in both studies, these values decrease to the similar values for the interval of the next hour. After an earthquake, the FFT method gives lower values of the period of excited waves than the method presented in this paper. Wave excitations with wave-periods of about 2 min obtained in the first 4 observed intervals are also visible in the study presented in \cite{nin20a} for intervals starting about 2 h before the earthquake and in the first hour after it. The post-earthquake wave-periods obtained in this study are also in agreement with those shown in \cite{ohy18} which indicate values from less than 10 s to a few hundred seconds.}

As previously, mentioned notable characteristics of our two-dimensional hybrid method include the simplification of complex spectra of detected oscillations that are composed of many overlapping peaks in Fourier periodograms, the enhancement of apparent spectral resolution through the spreading of peaks over the second dimension, and the establishment of the direction of changes in signal through correlation coefficients.
The normal { VLF signal amplitude} seen in control case one year before the earthquake shows more coherent topology of maps and oscillations below 111 s, whereas during the event of earthquake perturbations occur so that maps have more features off diagonal. This is in stark contrast to the observable oscillations of 120 and 130 seconds during earthquake.
Moreover, as oscillations of 140-147 s are documented in earthquake and quiet day time series data occurring 2 hours before the nominal earthquake occurrence, we predict this oscillation as usual behavior.
 
 Finally, we will take a moment reflecting about the ways in which our study could be expanded.
For the purpose of this investigation, we made use of the WWZ wavelet, which is defined in terms of trigonometric functions. However, in order to test for presence {of} non-sinusoidal oscillations, it is essential to use wavelets on a base that does not involve trigonometry. In light of this, we believe that Superlets will prove to be the most suitable option for the continuation of our research. 
Superlet is a spectral estimator enabling time-frequency super-resolution which  uses sets of wavelets with increasingly constrained bandwidth. These are combined geometrically in order to maintain the good temporal resolution of single wavelets and gain frequency resolution in upper bands. The normalization of wavelets in the set facilitates exploration of data with scale-free, fractal nature, containing oscillation packets that are self-similar across frequencies. Importantly, they can reveal fast transient oscillation events in single trials that may be hidden in the averaged time-frequency spectrum by other methods.

\section{Conclusion}

In this work, we explored the use of our 2D Hybrid technique for detecting oscillations in VLF signal amplitude time series in the time vicinity of the Kraljevo earthquake in Serbia in 2010 year. 
Furthermore, we demonstrated how the approach captures the difference between time series in the time vicinity of an earthquake and a control day one year earlier, which can be utilized to establish topology difference between certain ionosphere occurrences.

The cross correlation of scalograms of time series, which can be further integrated, is a crucial principle in our approach. This has two significant advantages.
First, the method simplifies complex spectra with numerous overlapped peaks in periodograms, i.e. increasing apparent spectral resolution by spreading peaks over the second dimension. For example, we were able to detect various oscillation patterns at various time series segments.

The second advantge is usage of correlation coefficients to determine the direction of signal changes. This enabled us to distinguish between cohesive topology of 2D maps during a calm day and more dispersed correlation clusters during Kraljevo earthquake.
We discovered that oscillations in the 120-130 s range appear 1 hour before the earthquake, continue to exist during the earthquake, and disappear 1 hour after the earthquake.
Fluctuations of the order 140-147 s occur 2 hours before the nominal start of the earthquake during the calm day and can be interpreted as normal oscillations in the VLF signal amplitude.
Finally, we briefly described how we could extend our research by using Superlets, which can reveal fast transient oscillation events that other types of wavelets may mask in averaged time-frequency scalograms.

% Example of a figure that spans the whole page width. The 
%%%%%%%%%%%%%%%%%%%%%%%%%%%%%%%%%%%%%%%%%%
\vspace{6pt} 

%%%%%%%%%%%%%%%%%%%%%%%%%%%%%%%%%%%%%%%%%%
%% optional
%\supplementary{The following supporting information can be downloaded at:  \linksupplementary{s1}, Figure S1: title; Table S1: title; Video S1: title.}


%%%%%%%%%%%%%%%%%%%%%%%%%%%%%%%%%%%%%%%%%%
\authorcontributions{
L. {\v C}.P.  conceptualized study, A.B.K. designed methodology, performed calculations, ploted figures and wrote the whole manuscript; A.N. collected data and  prepared parts related to VLF data and ionospheric observations; A.B.K., A.N., L.{\v C}.P. and M.R revised the manuscript. All authors have read and agreed to the published version of the manuscript.}

\funding{ABK and L{\v C}P acknowledge funding provided by University of Belgrade-Faculty of Mathematics  (the contract 451-03-68/2022-14/200104), through the grants by the Ministry of Education, Science, and Technological Development of the Republic of Serbia.
ABK and L{\v C}P thank the support by  Chinese Academy of Sciences President's International Fellowship Initiative (PIFI) for visiting scientist. AN acknowledges funding provided by the Institute of Physics Belgrade through the grant by the Ministry of Education, Science, and Technological Development of the Republic of Serbia.
{L{\v C}P} acknowledges funding provided by Astronomical Observatory (the contract 451-03-68/2022-14/ 200002), through the grants by the Ministry of Education, Science, and Technological Development of the Republic of Serbia.
 MR acknowledges funding provided by the Geographycal Institute "Jovan Cviji{\' c}" SASA through the grant by the Ministry of Education, Science, and Technological Development of the Republic of Serbia.
}

\institutionalreview{Not applicable.}

\informedconsent{Not applicable.}

\dataavailability{Requests for the VLF data used for analysis can be directed to AN.} 

\acknowledgments{The authors would like to express their gratitude to Reviewer 1 and Reviewer 2 for providing insightful comments that helped to strengthen the presentation of our work.}

\conflictsofinterest{ The authors declare no conflict of interest. The funders had no role in the design of the study; in the collection, analyses, or interpretation of data; in the writing of the manuscript; or in the decision to publish the~results.} 

%%%%%%%%%%%%%%%%%%%%%%%%%%%%%%%%%%%%%%%%%%
%% Optional
\sampleavailability{{\color{red}Not applicable.}}

%% Only for journal Encyclopedia
%\entrylink{The Link to this entry published on the encyclopedia platform.}

%\abbreviations{Abbreviations}{
%The following abbreviations are used in this manuscript:\\

%\noindent 
%\begin{tabular}{@{}ll}
%MDPI & Multidisciplinary Digital Publishing Institute\\
%DOAJ & Directory of open access journals\\
%TLA & Three letter acronym\\
%LD & Linear dichroism
%\end{tabular}
%}

%%%%%%%%%%%%%%%%%%%%%%%%%%%%%%%%%%%%%%%%%%

%%%%%%%%%%%%%%%%%%%%%%%%%%%%%%%%%%%%%%%%%%
\begin{adjustwidth}{-\extralength}{0cm}

\reftitle{References}


%\bibliography{your_external_BibTeX_file}

%=====================================
% References, variant B: internal bibliography
%=====================================

\appendixtitles{no} % Leave argument "no" if all appendix headings stay EMPTY (then no dot is printed after "Appendix A"). If the appendix sections contain a heading then change the argument to "yes".
\appendixstart
\appendix
\section[Pseudocode of 2D Hybrid method]{}

 In case readers would like to implement the 2D Hybrid method, we have provided implementation-agnostic pseudocode of our algorithm in Algorithm \ref{alg:cap}.
\begin{algorithm}
\caption{Pseudocode of 2D Hybrid method}\label{alg:cap}
\begin{algorithmic}
			\State Initilization:
			\State import wavelet module
			\State import Gaussian process module
			\State import time series 1 -$y_{1}$
			\State import time series 2 -$y_{2}$
		\State Initialize  parameters for   Gaussian process modeling if needed, otherwise set parameters as 0
		\State Compute Gaussian process model of $y_1$ and $y_2$ if parameters for modelling are nonzero 
		\Procedure{ 2D Hybrid map} {$y_{input1}, y_{input2}$}
			\State compute scalogram $\mathcal{S_1}$ of y\_input1
			\State  compute scalogram $\mathcal{S_2}$ of y\_input2
			\State  compute 2D correlation map as M=Cov($\mathcal{S_1}$, $\mathcal{S_2}$)
				\State Integrate 2D map along axis 1 $I_{1}=\sum_{i} M_{ij}$
			\State Integrate 2D map along axis 2 $I_{2}=\sum_{j} M_{ij}$
			\State  return ( $\mathcal{S_1}$,  $\mathcal{S_2}$, $I_1$, $I_2$)
			\EndProcedure
			\State  Calculate error of periods: determine FWHM of peaks in $I_1$ and $I_2$ with correlation larger than 0.5, and determine 25th and 75th quantiles, so that array of lower errors are 25th quantiles,  and array of upper errors are 75th quantiles
			\State  Calculate significance of period: set number of shuffling $N=100$
			\State  ynew=zeros(2,:)
		  \For  {$j \in (1,2)$}
			 \For {$i\in (N)$}
	          \State ynew[j,:,:]=random.shuffle($y_j$)
	          \EndFor
			\State  $\mathcal{S}[1,:,:]$,  $\mathcal{S}[2,:,:]$, $I[1,:,:]$, $I[2,:,:]=$procedure 2D Hybrid map (ynew[1,:,:], ynew[2,:,:])
			\EndFor
		     \State counter1=0, counter2=0
			\For {P1, P11 in zip(Peak($I_1$),Peak( $I[1,:,:]$))}
			\If {$P1>P11$}
			  \State $counter1=counter1+1$
			\EndIf
			\EndFor
			\For {P2, P22 in zip(Peak($I_2$),Peak( $I[2,:,:]$))}
			\If {$P2>P22$}
			  \State $counter2=counter2+1$
			\EndIf
			\EndFor
			\State $significance1=counter1/N, significance2=counter2/N$
			\State print P1, P2, lower error, upper error, $significance1, significance2$
\end{algorithmic}
\end{algorithm}

%%%%%%%%%%%%%%%%%%%%%%%%%%%%%%%%%%%%%%%%%%
%% for journal Sci
%\reviewreports{\\
%Reviewer 1 comments and authors’ response\\
%Reviewer 2 comments and authors’ response\\
%Reviewer 3 comments and authors’ response
%}
%%%%%%%%%%%%%%%%%%%%%%%%%%%%%%%%%%%%%%%%%%
\end{adjustwidth}
\end{document}